\documentclass[a4paper,12pt]{article}
\usepackage{jheppub}
\usepackage{amsfonts, amsmath}
\usepackage[english]{babel}
\usepackage{minibox}
\usepackage{amsmath, amssymb}
\usepackage{marginnote}
\usepackage{xcolor}
\usepackage{enumitem}
\usepackage[margin=45mm]{geometry}
\usepackage{setspace} 
\hypersetup{colorlinks,
  linkcolor=blue!60!black,
  citecolor=blue!50!black,
  urlcolor=cyan!60!black}

\title{Particle Cosmology: 1980--2000}

\author{Edward W. Kolb}

\affiliation{Kavli Institute for Cosmological Physics, \\
The University of Chicago\\  5640 South Ellis Avenue, Chicago, Illinois 60637, U.S.A.}

\emailAdd{Rocky.Kolb@uchicago.edu}

\abstract{
A new field, \emph{Particle Cosmology}, emerged at the interface of elementary particle physics and cosmology in the last two decades of the 20th century.  In this chapter of the \emph{Proceedings of the 4th International Symposium on the History of Particle Physics, the Standard Model and Beyond} I will review the development of Particle Cosmology in the period 1980-2000 centered on two events: a conference, \emph{Inner-Space/Outer-Space} held at Fermilab in May 1984 \cite{kolb1986innerspace} and the 1994 Snowmass Summer Study \emph{Particle and Nuclear Astrophysics in the Next Millennium} \cite{kolb1995particle}. 
}

\begin{document}
\maketitle

\section{Explorers of the particle cosmology frontier}\label{sec:intro}
\hspace{1.35in}
\begin{minipage}{0.70\textwidth}
\flushright \emph{Memory is like a sponge.  It soaks up material from other times, other places and leaks all over the moment in question.}\\
Ian McEwan, \emph{What We Can Know}\\
\end{minipage}
\bigskip

This contribution to the \emph{Proceedings of the 4th International Symposium on the History of Particle Physics, the Standard Model and Beyond} covers the emergence of particle cosmology as a frontier field, with emphasis on the years covered by the Symposium, 1980--2000.  While there were significant developments in other areas of cosmology, I will focus on the connections to elementary particle physics. 

 Historians of the frontier, especially historians writing about the settlement of the western United States, often describe the settlement of a frontier in terms of three waves: first the explorers, followed by the pioneers, and finally the settlers. Explorers are the first people to enter the unexplored territory.\footnote{As the history of the settlement of the U.S.\ frontier was originally written by Americans of European heritage, the fact that indigenous peoples already inhabited the lands is conveniently discounted.} Their goals were discovery and mapping. The \emph{Corps of Discovery Expedition}, commissioned in 1803 by U.S. President Thomas Jefferson to explore the recently-purchased western portion of the U.S., and led by Meriwether Lewis and William Clark, is an oft-used exemplar for exploration. Often in exploration some discoveries are unexpected, and just as often, expected discoveries are not realized.\footnote{An amusing example of an unrealized discovery of the Corps of Discovery is that in 1797 Jefferson wrote to George Rogers Clark (William Clark's brother) that the American west might have ``elephants and lions, if in that climate they could subsist; and the mammoths [mastodons] and megalonyxes may subsist there.'' }  Explorers typically did not remain in the territories they charted.  Then pioneers followed the explorers and took advantage of the discoveries and pathfinding of the explorers and built infrastructure by clearing the land, establishing outposts, and establishing the possibility of life in the frontier.  Unlike explorers, pioneers usually intended to remain on the frontier.   Finally, the settlers arrived and took advantage of the rudimentary infrastructure of the pioneers and provided for the long-term stability of the region.    

Scientists also often describe their study of nature in frontier terms. The most famous example of this is Vannevar Bush's influential 1945 report to U.S.\ President Harry S.\ Truman, \emph{Science, the Endless Frontier} \cite{bush1945science}.  I will follow the frontier imagery in discussing the emergence of particle cosmology as an integral part of particle physics.  

It is impossible to assign a definitive start date to a frontier field such as particle cosmology. It would be convenient for the purpose of this Symposium to date the beginning of the field of particle cosmology to 1980.  But of course there were prior explorations that we would recognize today as particle cosmology.  Some of the early explorations involved the role of neutrinos in cosmology; using cosmology to constrain the mass, lifetime, and number of neutrinos.  In 1966 Gershtein and Zel'dovich realized that the mass of neutrinos can be bound by demanding that the density of relic neutrinos not exceed the maximum matter density of the present Universe \cite{GershteinZeldovich1966}.   Their conclusion was that a neutrino mass must be less than 400 eV (they used a \emph{very} conservative estimate for the present mass density of the Universe).  Their calculation was refined in 1972 by Marx and Szalay ($m_\nu \leq 130$ eV) \cite{Marx1972} and by Cowsik and McClelland ($m_\nu\leq 8$ eV) \cite{Cowsik:1972mw}. After the 1975 discovery of the third charged lepton \cite{Perl:1975bf}, the $\tau$ lepton, it was widely assumed that there was an associated third-generation neutrino.  The question arose whether there were additional lepton generations with associated neutrinos awaiting discovery.  In 1977 Steigman, Schramm, and Gunn made use of Big-Bang Nucleosynthesis to limit the number of light neutrinos so as not to overproduce primordial helium  They found there could be no more than five light neutrinos (i.e., no more than two additional neutrinos awaiting discovery) \cite{Steigman:1977vz}.  Also in 1977, five groups independently realized that the neutrino mass limit could be evaded if neutrinos were sufficiently massive, say larger than a few GeV.  In the book  \emph{Cosmology's Century} \cite{peebles2020cosmology}, P.\ J.\ E. Peebles credits near simultaneous papers by Hut \cite{Hut:1977zn}, Lee and Weinberg \cite{Lee:1977qn}, Sato and Kobayashi \cite{Sato:1977wd}, Dicus, Kolb, and Teplitz \cite{Dicus:1977nn}, and Vysotskii, Dolgov, and Zel'dovich \cite{Vysotskii1977} for the introduction of GeV mass neutrinos as the Cold Dark Matter (CDM) prototype.

Axions as a solution to the strong CP problem were proposed by Wilczek \cite{Wilczek:1978gs} and Weinberg \cite{Weinberg:1977ma} in 1978.  It was soon realized limits on the energy carried off from stars (like the sun) by the emission of light, weakly interacting particles could place limits on the properties of axions \cite{Dicus:1978fp}.  Limits on stellar energy loss by emission of weakly-interacting particles continues to be an active area of research.

In 1974 Howard Georgi and Sheldon Glashow proposed a Grand Unified Theory (GUT) based on the group SU(5) \cite{Georgi:1974sy}.  In GUTs, baryon number violation is a consequence of having quarks and leptons in the same multiplets of the gauge group.  In SU(5), baryon-number violating reactions are mediated by the exchange of massive vectors ($X$ and $Y$ bosons) and scalars.  The masses of these new baryon-number violating gauge and Higgs bosons were expected to be very high, far beyond the range of terrestrial accelerators, so it was widely realized that the early Universe provides a laboratory (perhaps the only laboratory) to study unified theories.  Baryon-number violation in GUTs provides one of the Sakharov ingredients \cite{Sakharov:1967dj} needed to generate the baryon number of the Universe.  The possibility that the decay of $X$ bosons generates the baryon asymmetry drew many particle physicists to cosmology \cite{Weinberg:1979gw,Yoshimura:1978jy,Dimopoulos:1978kv,Toussaint:1978br,Ellis:1978xg}: they are known as the Gen-$X$-boson cosmologists.  After the non-observation of proton decay, particle physicists turned to the idea that baryogenesis occurred at the electroweak scale (roughly the mass of the electroweak $Z$ boson) through the effect of sphalerons \cite{Klinkhamer:1984rz}.  This popular idea brought into the study of cosmology Gen-$Z$-boson particle physicists, but that is a subject for a later chapter.

These examples, limits on neutrino and axion properties and baryogenesis are examples of the exploration phase of cosmology by particle physicists. 

Prior to 1970, cosmology was a field studied by astronomers, astrophysicists, and general relativists, but not by particle physicists.  Nuclear astrophysics was a recognized field, but particle cosmology or particle astrophysics was not.  But in the 1970s that started to change.  By the start of the time frame for this Symposium (1980), the exploration phase of \emph{particle} cosmology was at an end. Prior to 1980 the particle cosmology frontier had been explored by well-known particle physicists such as Dimopoulos, Ellis, Gaillard, Nanopoulos, Susskind, Weinberg, Wilczek, Zee, and others.  Their early explorations mapped out how particle physics could use the early Universe as a laboratory, and how cosmologists might make use of advances in particle physics to understand the evolution and present structure of the Universe.

\section{Inner Space/Outer Space, the Interface between Cosmology and Particle Physics: Pioneers of the particle cosmology frontier}\label{sec:ISOS}

\emph{Inner Space/Outer Space, the Interface between Particle Physics and Cosmology} (IS/OS) was a conference held at Fermi National Accelerator Laboratory in May, 1984. The conference was organized by the newly-formed ``astrophysics'' group at Fermilab. The fact that an astrophysics group was part of a high-energy accelerator laboratory is a clear indication that astrophysics and cosmology was starting to become a recognized part of particle physics.

The genesis of the Fermilab Astrophysics Group was a conversation during a hike in the Dolomites by Fermilab Director Leon Lederman and University of Chicago astrophysicist David Schramm.  Schramm suggested to Lederman that particle astrophysics and cosmology could have a home at Fermilab.  This inspired Lederman to start an Astrophysics Group with NASA support.  I don't believe this was ever officially approved by the Department of Energy: it seems that in the early 1980s laboratory directors had more discretion than they do now.

In the fall of 1983, Michael Turner and I arrived at Fermilab to lead the group, along with the first wave of postdocs: David Lindley, Keith Olive, and David Seckel. By the year 2000, about 35 postdocs had been part of the astrophysics group, along with many visitors.  Our goal was to make Fermilab, along with the University of Chicago, the home on the prairie for pioneers, and later settlers, of the particle cosmology frontier.

We (Kolb, Lindley, Olive, Seckel, and Turner) decided to organize an international conference to advertise the group, and for the first time to bring together cosmologists, particle theorists, particle experimentalists, astrophysicists, relativists, cosmic-ray physicists, and low-temperature physicists under one roof and explore common interests.   I recall at the time having some trepidation that scientists from the targeted communities would not attend.  But the worry turned out to be unfounded since the conference had over 200 participants.  

The attendees included people well-known in their areas, but perhaps not widely known to those in other areas. It is instructive to list some names.  Among particle physicists were Larry Abbott, Steve Barr, Bill Bardeen, Bj Bjorken, Marty Einhorn, Steve Ellis, Peter Freund, Haim Goldberg, Jeff Harvey, Boris Kayser, Paul Langacker, Joe Polchinski, John Preskill, Chris Quigg, Mark Srednicki, Erick Weinberg, Steve Weinberg, and Motrohiko Yoshimura.  

Cosmology theorists included John Bahcall, Dick Bond, Alan Guth, Richard Gott, Rich Holman, Nick Kaiser, Katsuhiko Sato, Pierre Sikivie, Joe Silk, Gary Steigman, Paul Steinhardt, Alex Szalay, Alex Vilenkin, and Simon White.  

Astro/Cosmo observers also attended, including Marc Davis, Dave De Young, Jim Gunn, John Huchra, Bernard Pagel, Paul Richards, and David Wilkinson.

Relativists were also represented: Bernard Carr, Jim Hartle, Ron Kantowski, Leonard Parker, Malcolm Perry, Remo Ruffini, Bob Wald, and Cliff Will, among others.

Experimental particle physics were also there: Barry Barish, Henry Frisch, Joe Incandela, Leon Lederman, Adrian Melissinos, Frank Sciulli, and Peter Trower.

And to ensure that the conference was magical, we inited a professional magician, The Amazing Randi, to be the after-dinner speaker.  This was perhaps the first and only physics meeting to feature a professional magician.

\emph{Inner Space/Outer Space} is the best snapshot of the field in the mid-1980s, both in terms of the range of subjects and the growing confidence of particle physicists and astronomers that cosmology could be used as a tool of particle physics, and conversely, particle physics could make use of cosmology and astronomy.

The scientific program of the conference had ninety presentations in nine subject areas encompassing the overlap between particle physics and cosmology circa 1984: 
\begin{enumerate}[noitemsep]
\item Standard models of particle physics and cosmology (10 presentations).
\item Microwave background radiation (8 presentations).
\item Origin and evolution of large-scale structure (19 presentations).
\item Inflationary Universe (10 presentations).
\item Massive magnetic monopoles (10 presentations).
\item Supersymmetry, supergravity, and quantum gravity (9 presentations).
\item Cosmological constraints on particle physics (13 presentations).
\item Cosmology in extra dimensions (9 presentations)
\item Directions and connections in particle physics and cosmology (2 presentations).
\end{enumerate}

It is instructive to dive into a few of the subject areas to illustrate how the particle physics/cosmology interface rapidly developed in the first half of the 1980s.

In 1981 and 1982 cosmic inflation was proposed by Starobinsky, Guth, and Sato \cite{Starobinsky:1980te, Guth:1981vz,Sato:1981ds}, and shortly thereafter developed by Linde \cite{Linde:1983gd} and Albrecht and Steinhardt \cite{Albrecht:1982wi}.   The excitement of the early days of the inflationary Universe is illustrated by the number of talks about inflation:
\begin{enumerate}[noitemsep]
\item The new inflationary Universe, A. Guth
\item Supersymmetric inflation, B. Ovrut \& P. Steinhardt
\item Inflation in $N = 1$ supergravity, M. Srednicki
\item Fluctuations in cosmological models, R. Brandenberger
\item Behavior of the Higgs field in new inflationary Universe, A. Guth \& S.-Y. Pi
\item Evolution of classical Higgs field and fate of inflationary Universe, K. Sato \& H. Kodama
\item Conditions for formation of bubble universes, J. R. Gott
\item Inflation in wall-dominated Universe, D. Seckel
\end{enumerate}
Note that supersymmetry leaked into cosmological models of inflation.

\emph{Inner Space/Outer Space} predated the First Superstring Revolution of 1984 \cite{Green:1984sg} by just a few months.  But the role of the extra dimensionsin the form of Kaluza-Klein theories garnished quite a lot of attention. Talks on cosmology in extra dimensions included:
\begin{enumerate}[noitemsep]
\item Physics in higher dimensions, S. Weinberg
\item Kaluza-Klein cosmology: Techniques for Quantum Compactification, M. Rubin
\item High-temperature quantum effects in Kaluza-Klein cosmology, M. Yoshimura
\item More dimensions-less inflation, D. Lindley
\item Inflation from extra dimensions, S. Barr
\item Massless fermions in Kaluza-Klein theories, G. Venturi
\item Reduced gauge group of symmetric Kaluza-Klein space, R. Kantowski 
\end{enumerate}

By 1984 supersymmetry was gaining in popularity among particle physicists, but it is fair to say that most astronomers had never heard of photinos, axinos, or gravitinos before \emph{Inner Space/Outer Space}.  There were very exciting sessions on Supersymmetry, Supergravity, and Quantum Gravity, with talks including:
\begin{enumerate}[noitemsep]
\item Supersymmetry and supergravity, J. Polchinski
\item Supersymmetric relics from the big bang, J. Hagelin
\item Is spontaneously broken supersymmetry restored at high temperature, M. Kulab
\item The photino, the axino, and the gravitino in cosmology,   J. Kim
\item Higgs masses and supersymmetry, R. Flores \& M. Sher
\item Initial conditions, J. Hartle
\item Path integral quantum cosmology, B. Berger
\end{enumerate}

The Nobel Prize winning discovery of anisotropies in the cosmic microwave background (CMB) \cite{Smoot:1992lz} would not come until nearly a decade after \emph{Inner Space/Outer Space}, but it was clear that the search was well underway.  There were three presentations on upper limits to anistropies in the CMB, by Wilkinson, Uson, and Partridge. Joe Silk spoke of the implications of CMB measurements for galaxy formation and the Bond and Efstathiou paper discussed how CMB anisotropies constrain dark matter.

The last subject area I would like to highlight is \emph{Other Particle-Cosmology Topics}.  Talks in this session included:
\begin{enumerate}[noitemsep]
\item Axions in astrophysics and cosmology, P. Sikivie
\item Role of axions in neutron star cooling, N. Iwamoto
\item Mirror fermions and cosmology, G. Senjanović
\item Review of neutrinos and neutrino mass from particle physics experiments, F. Sciulli
\item New and old accelerators, Bj Bjorken
\item Axions, neutrinos, and strings, F. Stecker
\item Present status of grand unification and proton decay, P. Langacker
\end{enumerate}
The idea that axions from the early Universe could be dark matter was proposed by three independent groups; Preskill, Wise, and Wilczek \cite{Preskill:1982cy}, Abbott and Sikivie \cite{Abbott:1982af}, and Dine and Fischler \cite{Dine:1982det} just a year before \emph{Inner Space/Outer Space}.  Axion cosmology would grow in importance and is still a growing subject at the particle/cosmology interface.

The development of today's Cold Dark Matter (CDM) theory for the evolution of structure in the Universe is reviewed at length in Peebles book \cite{peebles2020cosmology}.  Here I will only note that by 1984 particle physics was providing particle candidates for CDM (inos, axions, etc.), but hot dark matter still had many adherents. Marc Davis reviewed the status of what was then considered to be large-scale galaxy surveys, but without speculating on the nature of dark matter.  Simon White opened his contribution by stating ``Cross fertilization between particle physics and cosmology has introduced a number of new elements into the old problem of understanding the large-scale structure of the Universe.  At the simplest level particle physics has come up with a number of elementary particle candidates for the dark matter...the most plausible from the point of view of particle physics, was a massive neutrino...''  Massive (light) neutrinos would be hot dark matter, rather than cold dark matter.  Although White considered neutrinos to be the ``most plausible'' dark matter candidate, comparison of numerical simulations of structure formation with galaxy redshift surveys had begun to favor cold dark matter. 

From \emph{Inner Space/Outer Space} it is clear that by 1984 much of the now current areas of research in Particle Cosmology were in place: cold dark matter, limits on the properties of elementary particles, the connection between inflation and temperature/density perturbations, and the search for temperature fluctuations in the CMB.

Some of the areas covered at the meeting did not pan out: e.g., magnetic monopoles, Kaluza-Klein, proton decay and GUTs, cosmic strings as seeds of structure formation, have faded in relevance.

Another future Nobel Prize winning cosmological result was the 1998 discovery of the acceleration of the Universe \cite{Riess:1998cb,Perlmutter:1998np}.  This discovery was unanticipated at the time of \emph{Inner Space/Outer Space}.  In 1984 theorists thought that the Universe was spatially flat with the mass-energy density dominated by dark matter, while observers stubbornly insisted that data from large-scale surveys implied the matter density could not be enough to make a spatially flat Universe.  By the end of the 1990s it was clear that theorists were right that the Universe was spatially flat and observers were correct that the matter density was too low to account for this.  

If \emph{Inner Space/Outer Space} indicated the settlement of the particle cosmology frontier, it is appropriate to ask, why there, why then? With the establishment of the Theoretical Astrophysics Group at Fermilab, I believe it was natural to have the meeting there. In the conclusion section I will address the ``why then'' question by asking if such a meeting was conceivable ten years earlier, say in May 1974?

\section{Particle and Nuclear Astrophysics in the Next Millennium, the 1994 Snowmass Summer Study:  Settlers on the particle cosmology frontier}\label{sec:Snowmas}

If by the mid-1980s particle cosmology transitioned from the exploration phase to the pioneer phase, by the time of the \emph{1994 Snowmass Summer Study} the pioneer phase of particle cosmology had ended, the particle cosmology field was settled, and it would remain a vibrant area of particle physics. 

The U.S.\ particle physics community periodically convenes a series of planning meetings organized under the auspices of the Division of Particles and Fields of the American Physical Society.  The first Snowmass Summer Study was held in Snowmass, Colorado in 1982, and continued to be held in Snowmass roughly every other year before developing into a peripatetic meeting in the last decade.  The purpose of these highly influential planning meetings was to identify the most crucial questions in the field and how they might be addressed.  

Many of us in the particle cosmology community felt that the Snowmass process had not been representative of cosmology and astrophysics.  For instance, Inspire lists 133 contributions to Snowmass 1990, but with only 4 related to cosmology or astrophysics.  This led Roberto Peccei and myself to propose a Snowmass Summer Study devoted to particle and nuclear astrophysics.  In keeping with the Snowmass tradition of long-term planning, we included ``in the Next Millennium'' in the title (confident particle cosmology was here to stay).

Snowmass 1994 was a two-week workshop (June 29-July 14 1994) held in Snowmass under the auspices of The Astrophysics Division, the Division of Nuclear Physics, and the Division of Particles and Fields of the American Physical Society.  Roberto and I recruited five conveners: Jim Cronin, Wick Haxton, Bernard Sadoulet, Paul Steinhardt, and Kip Thorne.  Cronin and Sadoulet were recovering particle experimentalists now working on experimental astrophysics, Haxton is a nuclear physicist, Steinhardt a cosmologist, and Thorne a relativist.

The Introduction to the Proceedings was written by the organizers and conveners.  Rereading the Introduction thirty years after it was written, two statements jump out at me.  The first is how the ``organizational and computational talents of particle physicists emboldened astronomers [to undertake large projects].''  The influence of particle physics in astronomy eventually led to large ground-based astronomy/cosmology projects such as the Sloan Digital Sky Survey \cite{York2000}, and the Pierre Auger Observatory \cite{PierreAuger:2015}.  The second statement is ``[It is a] sad testament to the times that the potential for new discoveries does not seem to be limited by a lack of ideas, technology, or proposals, but by fiscal realities and the artificially constructed barriers of the existing science policy framework.''  Perhaps the same statement could be made today.

Snowmass 1994 had over 450 participants.  Clearly by 2024 the particle cosmology/astrophysics frontier was settled. There were 20 working groups and 5 supergroups.  The supergroups, organizers, and some of the topics were
\begin{enumerate}[noitemsep]
\item \emph{Neutrinos} (Wick Haxton) mass, oscillations, solar neutrinos, neutrino astrophysics.
\item \emph{Cosmic Rays} (Jim Cronin \& Jonathan Ormes)  Space-based $\gamma$-ray, ground-based $\gamma$-ray, over the knee cosmic rays, highest-energy cosmic rays.
\item \emph{Gravitational Phenomena} (Kip Thorne) Black holes and cosmology, black hole astrophysics, gravity waves.
\item \emph{Low-Background Experiments} (Bernard Sadoulet) proton decay, dark matter, other underground experiments.
\item \emph{Cosmology} (Paul Steinhartd) cosmological models, baryogenesis, nucleosynthesis, cosmic background radiation, structure formation (simulation and phenomenology), and observational cosmology.
\end{enumerate}

In all, there were 60 papers in the Snowmass proceedings. Just as the proceedings of \emph{Inner Space/Outer Space} was a window into the status of the field in 1984, the Proceedings of Snowmass 1994 provides a snapshot of the field a decade later. Comparison of the two proceedings can be used to study what changed in the decade between the meetings.
\begin{enumerate}[noitemsep]
\item Perhaps the biggest change were the COBE discoveries of temperature anisotropies in the CMB \cite{Smoot:1992lz} and precision measurements of its blackbody spectrum \cite{Mather1990}.  The COBE results strengthened evidence for inflation and the theory of formation of large-scale structure by the growth  of primordial seeds through gravitational instability. At the meeting George Smoot summarized the status of CMB anisotropy experiments. 
\item Physicists took the idea of weakly-interacting particle dark matter seriously enough to plan the first generation of dark-matter direct-detection experiments.
\item Laboratory axion search experiments were starting.
\item High-energy cosmic rays became part of the conversation.  Cronin presented plans for a giant air-shower array, which later became the Pierre Auger Observatory.
\item The combination of advances in large-scale structure observations and numerical simulations strengthened the idea that cold dark matter and inflation-produced seed of structure CDM emerged as the most promising model.  Hot dark matter (neutrinos) were now disfavored.
\item Francis Halzen presented the case for a square-kilometer scale neutrino detector that eventually became the IceCube Neutrino Observatory.
\item Kip Thorne spoke about the the Laser Interferometer Gravitational Wave Observatory (LIGO). Gravitational waves and black-hole astrophysics was just beginning to attract widespread attention outside the GR community.  Of course it wasn't until 2016 that LIGO first detected gravitational waves from a black-hole inspiral \cite{Abbott:2016bln}.
\item String theory was only lightly represented.
\item Neutrino oscillations (solar, atmospheric, accelerator) were more prominent than in the 1984 meeting. 
\end{enumerate}

The fact that 450 physicists, astronomers, and astrophysicists came together for two weeks in the mountains of Colorado was evidence that the particle cosmology frontier was settled.  By 1994 particle cosmology was an important component of particle physics.  This can be seen in the report of the U.S. 2023 Particle Physics Project Prioritization Panel (P5) \cite{P5Report2023} report. 

Of course discoveries continued.   The Hubble Space Telescope Key Project \cite{Freedman2001} pinned down the value of the Hubble constant to an uncertainty of 10\%.  The BOOMERanG \cite{deBernardis:2000sgs} balloon flights from Antarctica (along with other observations) observing CMB anisotropies demonstrated that we do indeed live in a spatially flat Universe. The suprising 1998 discovery of the acceleration of the Universe \cite{Riess:1998cb,Perlmutter:1998np} reconciled inflationary predictions and CMB observations of a spatially flat Universe with astronomical evidence that the mass density of dark matter was insufficient to provide flatness; most of the mass-density of the universe is in something that resembles a cosmological constant $\Lambda$.  A \emph{Cosmological Standard Model}, $\Lambda$CDM was in place by the year 2000, based upon ideas developed in the years 1980-2000.

Naturally, the settlement of the frontier continues today as questions still remain. For instance, despite heroic experimental efforts and many compelling theoretical ideas, the nature of dark matter remains a mystery.  We still do not have a compelling particle physics model for inflation.  And, in my opinion there is no understanding of the smallness of the present cosmological constant.  

\newpage

\section{Conclusions}\label{sec:Conclusion}
\hspace{1.5in}
\begin{minipage}{0.75\textwidth} 
\emph{Memory believes before knowing remembers.\\ Believes longer than recollects, \\ longer than knowing even wonders.}\\
\hspace*{48pt}William Faulkner, \emph{Light in August}\\
\end{minipage}

A frontier field, Particle Cosmology, emerged in the 1980s and 1990s. Originally on the fringes, cosmology is now firmly embedded in particle physics.  Ideas from particle physics drove much of modern cosmology, both theoretical and observational.

I have used the frontier imagery to trace the exploration era prior to 1980, to the pioneer period of the mid-1980s, finally to the settler epoch starting in the mid-1990s.  I used the proceedings of the 1984 conference \emph{Inner Space/Outer Space} to examine the status of particle cosmology in the mid 1980s, and the Proceedings of \emph{Snowmass 1994} to illustrate the status of particle cosmology in the mid-1990s.

I believe that Chicagoland (Fermilab and the University of Chicago) played a central role in the development of the field, especially in 1980s.  That is not to say that other people and institutions were not influential.  For instance, the theory group at CERN (Ellis, Nanopoulas, Gaillard, et al.) were very active in the field.  But Fermilab had the only group dedicated to particle cosmology.

If \emph{Inner Space/Outer Space} in 1984 was the ``coming-out'' party for particle cosmology, it is interesting to return to the question of ``why there,'' and ``why then.'' I remarked previously that ``why there'' is because Fermilab, a Department of Energy accelerator laboratory, started the first group dedicated to particle cosmology and astrophysics.  ``Why then'' is perhaps more difficult to answer, but I don't believe the time was ripe a decade earlier, say in 1974, to have \emph{Inner Space/Outer Space: The Interface Between Cosmology and Particle Physics}.  

In 1974, particle physicists wondering if the early Universe could be employed as a laboratory probably would have learned about cosmology through Steven Weinberg's book \emph{Gravitation and Cosmology, Principles and Applications of the General Theory of Relativity}, published in 1972  \cite{weinberg1972gravitation}.  After consulting Weinberg's book they might have thought that the big bang was \emph{not} a useful tool to study high-energy physics.  In Section 11 (The Very Early Universe) of Chapter 15 (Cosmology: The Standard Model), Weinberg writes ``However, if we look back a little further, into the first 0.0001 sec of cosmic history, when the temperature was above $10^{12\circ}$K [$8.6\times10^{-2}$\,GeV], we encounter theoretical problems of a difficulty beyond the range of modern statistical mechanics.''  Weinberg goes on to describe two pictures, the  \emph{composite particle model} and the \emph{elementary particle model}.  In the composite particle picture there would be a maximum temperature to the Universe, which on the basis of the Hagedorn \cite{Hagedorn:1965} and Veneziano \cite{Veneziano:1968yb} models Weinberg estimated to be of order $1.7\times10^{12\circ}$K [$1.5\times10^{-1}$\,GeV]. In the elementary particle picture, Weinberg supposes that there would be an ideal gas of ``elementary particles--say photons, leptons, `quarks,'\footnote{Note that Weinberg places quarks in quotation--the book was published two years before the discovery of the $J/\Phi$ \cite{Augustin:1974HO,Aubert:1974PC}, which lead to the widespread acceptance of the quark model.} and their antiparticles.'' Just one year after Weinberg's book was published, Gross and Wilczek \cite{Gross:1973id} and Politzer \cite{Politzer:1973fx} discovered asymptotic freedom. This was shortly followed in 1975 by Collins and Perry \cite{Collins:1975sx} using asymptotic freedom as support for the ``elementary particle'' picture, pointing out ``The quark model implies that superdense matter … consists of quarks rather than hadrons.  Bjorken scaling implies quarks interact weakly.  An asymptotically free gauge theory allows realistic calculations taking full account of strong interactions.''  The ``hadron barrier'' was broken.  Perhaps in 1974 particle physics had not developed to the point where it could be applied to the early Universe.  Weinberg's final chapter (Cosmology: Other Models) briefly discusses models with a cosmological constant (prescient), revisits the Steady State Model (why?), and ends with a discussion of ``Models with a Varying Constant of Gravitation'' (strange in 1974).  This speaks to a lack of confidence that cosmology could be extended to study conditions of energies above the hadronic scale.

An astrophysicist interested in the early Universe probably would have turned to Peebles's book \emph{Physical Cosmology} \cite{peebles1971}, published in 1971.  Peebles goes no earlier in the history of the Universe than big-bang nucleosynthesis, and does not speculate on what we would now call ``the early Universe.''  

If using Weinberg's or Peebles's books, one would be justified in believing that the early Universe (say prior to big-bang nucleosynthesis) was \emph{terra incognita} and not a proper place for the study of particle physics.

Finally, in 1974 someone asking what cosmology is all about might have read the \emph{Physics Today} article by Allan Sandage, \emph{Cosmology: A search for two numbers} (the Hubble constant and the deceleration parameter) \cite{Sandage1970} and concluded that if those two numbers is all cosmology is about, it would be of no use to particle physicists.

I conclude that a conference dedicated to the interface of particle physics and cosmology would not have been feasible in 1974, and unlikely much before 1984.

Here I have presented my personal memories of the 1980-2000 period of particle cosmology.  Given time and space limitations there are many important discoveries, papers, and players that were not mentioned.  Apologies to those omitted. 

It was (and still is) terribly exciting to have a ringside seat (and sometimes a seat \emph{in the ring}) in the convergence of Inner Space and Outer Space.

\bibliographystyle{JHEP} 
\bibliography{bibliography_file} 

\end{document}